\documentclass[twocolumn,  tighten,  twocolappendix]{aastex631} 

\shortauthors{Yoon et al. (2024)}

\begin{document}

\title{Shell-type Tidal Features Are More Frequently Detected in Slowly Rotating Early-type Galaxies than Stream- and Tail-type Features}

\email{yyoon@kasi.re.kr}

\author[0000-0003-0134-8968]{Yongmin Yoon}
\affiliation{Korea Astronomy and Space Science Institute (KASI), 776 Daedeokdae-ro, Yuseong-gu, Daejeon 34055, Republic of Korea}

\author[0000-0002-9434-5936]{Jongwan Ko}
\affiliation{Korea Astronomy and Space Science Institute (KASI), 776 Daedeokdae-ro, Yuseong-gu, Daejeon 34055, Republic of Korea}
\affiliation{University of Science and Technology, Gajeong-ro, Daejeon 34113, Republic of Korea}

\author[0000-0002-3043-2555]{Haeun Chung}
\affiliation{University of Arizona, Steward Observatory, 933 N Cherry Avenue, Tucson, AZ 85721, USA}

\author[0000-0002-7762-7712]{Woowon Byun}
\affiliation{Korea Astronomy and Space Science Institute (KASI), 776 Daedeokdae-ro, Yuseong-gu, Daejeon 34055, Republic of Korea}

\author[0000-0001-9544-7021]{Kyungwon Chun}
\affiliation{Korea Astronomy and Space Science Institute (KASI), 776 Daedeokdae-ro, Yuseong-gu, Daejeon 34055, Republic of Korea}

\begin{abstract}
To enhance our understanding of the impact of galaxy mergers on the kinematics of early-type galaxies (ETGs), we examine differences in specific stellar angular momentum within the half-light radius ($\lambda_{R_e}$) among ETGs with different types of tidal features and those without such features. This is accomplished by categorizing tidal features, which serve as direct evidence of recent mergers, into shells, streams, and tails, through deep images from the DESI Legacy Survey, and by using MaNGA data for the analysis of the kinematics of 1244 ETGs at $z<0.055$. We find that ETGs with tidal features typically have reduced $\lambda_{R_e}$ values that are lower by 0.12 dex than ETGs without tidal features. ETGs with shells contribute most to the reduction in $\lambda_{R_e}$. Consequently, nearly half of ETGs with shells are classified as slow rotators, a fraction that is more than twice as high as that of ETGs with tails or streams, and over three times higher than that of ETGs without tidal features. These trends generally remain valid even when ETGs are divided into several mass bins. Our findings support the idea that radial mergers, which are more effective at reducing $\lambda_{R_e}$ than circular mergers, are more closely associated with the formation of shells rather than streams or tails. The detection of shells in slightly more massive ETGs compared to streams and tails may be attributed to the fact that massive satellite galaxies are more likely to be accreted through radial orbits, due to the nature of dynamical friction.
\end{abstract}

\section{Introduction}\label{sec:intro}
Early-type galaxies (ETGs) are in the mature stage of the cosmic evolution of galaxies, predominantly composed of old stellar populations. The low star formation rate of ETGs is a consequence of the depletion of cold gas in these systems. As a result, ETGs generally exhibit red colors ($g-r\gtrsim0.7$) in the optical bands \citep{Gallazzi2006,Graves2009,Schawinski2014,Lacerna2020}. ETGs tend to possess smoother and simpler structures compared to late-type galaxies, which often exhibit spiral arms with intricate features and blobs of highly active star-forming regions \citep{Nair2010}. The stars in ETGs are distributed with a higher light concentration toward the center compared to those in late-type galaxies \citep{Park2005,Choi2010}.  Accordingly, the modeling of ETG surface brightness profiles typically employs centrally concentrated light profiles, such as the de Vaucouleurs profile or the S{\'e}rsic profiles with high S{\'e}rsic indices of $n\gtrsim3$ \citep{Blanton2009,Huertas-Company2013}.  

Beyond uncovering these photometrically derived ETG properties, a comprehensive understanding of the kinematic properties of ETGs has been attained through large surveys based on integral field unit (IFU) spectroscopy \citep{Bacon2001,Bershady2010,Cappellari2011,Sanchez2012,Ma2014,Yoon2021,Sanchez2016}. IFU surveys have provided new insights into ETGs. For instance, studies utilizing IFU survey data have revealed that a large portion of ETGs, including those with round and nondisky shapes, possess stellar components that exhibit significant rotation similar to disks in late-type galaxies \citep{Cappellari2016,Graham2018}. Thus, a multitude of studies utilized a new classification system that categorizes ETGs into either slow or fast rotators based on resolved kinematics \citep{Emsellem2007,Emsellem2011,Jesseit2009,Khochfar2011,Cappellari2016,Graham2018}. Indeed, this is increasingly being regarded as a more physically meaningful classification, replacing the conventional categorization into ellipticals and lenticulars, which strongly depends on line-of-sight inclinations \citep{Emsellem2007,Cappellari2011,Cappellari2016}.

In the standard $\Lambda$ cold dark matter universe, galaxy mergers play a crucial role in the formation and evolution of ETGs \citep{Baugh1996,Christlein2004,DeLucia2006,DeLucia2007,Wilman2013,Yoon2017}, shaping the fundamental properties of ETGs. This is particularly significant for massive ETGs with $M_\mathrm{star}>10^{10.5}\,M_{\odot}$, for which the contribution of ex-situ sources to mass assembly is considerable \citep{Dubois2016,Davison2020}. For example, galaxy mergers have the potential to generate red and quiescent remnants in the end that do not actively produce young stellar populations \citep{Springel2005,Hopkins2008,Brennan2015}, owing to the rapid depletion of available cold gas through intense star formation during the merger processes \citep{Hernquist1989,Mihos1996,Springel2005}. Strong feedback effects from active galactic nuclei (AGNs) triggered by the merger processes may also have the capability to quench the merger remnants \citep{Hopkins2005,Hopkins2008,Springel2005}. In addition, the concentrated steep light profiles, commonly observed in typical ETGs, can stem from mergers \citep{Barnes1988,Naab2006,Hilz2013}.

Likewise, galaxy mergers are expected to impact the stellar kinematics of ETGs, given that more massive ETGs, which more predominantly form and grow through mergers \citep{DeLucia2006,DeLucia2007,Yoon2017}, typically have lower specific stellar angular momentum \citep{Emsellem2007,Graham2018}. Numerical simulations have been employed in previous studies to elucidate the influence of galaxy mergers on the stellar kinematics of merger remnants. For instance, the mass ratios and merger orbits of merger progenitors play pivotal roles in determining the kinematics of merger remnants, according to \citet{Jesseit2009}, \citet{Bois2011} and \citet{Martin2018}. The simulation of \citet{Choi2017} demonstrated that galaxy mergers tend to statistically reduce rotation speeds, especially in massive galaxies. They also observed that frequent minor mergers exert significant cumulative effects on the kinematics of merger remnants. Similarly, \citet{Lagos2018a} and \citet{Schulze2018} also discovered a reduction in stellar rotations as a result of galaxy mergers in their simulations. In addition, the simulation of \citet{Choi2018} suggested that galaxy mergers are the dominant driver of the spin reduction for central ETGs in dense regions.

In contrast to simulation studies, conducting direct observational studies on the effect of galaxy mergers on the properties of ETGs is relatively challenging. This is due to the fundamental limitation that we can only observe a snapshot of the universe, rather than the continuous flow of cosmic time. Nevertheless, we can mitigate this limitation by using tidal features. Tidal features are stellar debris generated by galaxy mergers \citep{Toomre1972,Quinn1984,Barnes1988,Hernquist1992,Feldmann2008}, which are generally fainter than the main bodies of galaxies, necessitating deep images for their detection and examination. Hence, tidal features such as tidal tails, streams, and shells serve as the most direct observational indicators of recent mergers, offering a means to examine the impact of mergers on the photometric \citep{Schweizer1992,Tal2009,Schawinski2010,Kaviraj2011,Hong2015,YL2020} and kinematic \citep{Krajnovic2011,Duc2015,Oh2016,Yoon2022,Bilek2023} properties of ETGs.

For example, it has been observed that ETGs displaying blue optical colors are more prone to possess tidal features or morphological disturbances \citep{Schweizer1992,Tal2009,Schawinski2010,Kaviraj2011}, suggesting a potential association between young stellar populations in ETGs and recent merger events. In a study by \citet{Hong2015}, it was found that nearly half of luminous AGN hosts, primarily ETGs, exhibit tidal features, which is in contrast to the lower fraction observed in typical ETGs. This indicates that luminous AGNs in ETGs are likely to be activated by recent merger events. Combining tidal features identified in deep images with IFU spectroscopic data, \citet{Yoon2023} uncovered in detail that ETGs that have undergone recent mergers have different stellar population profiles compared to their counterparts that have not experienced recent mergers. Recently, \citet{Yoon2022} studied the impact of galaxy mergers on stellar kinematics of ETGs, using 167 ETGs in the Mapping Nearby Galaxies at Apache Point Observatory (MaNGA; \citealt{Bundy2015,Drory2015,Yan2016,Wake2017}) IFU data that are in the Stripe 82 region of the Sloan Digital Sky Survey (SDSS). The main discovery of \citet{Yoon2022} is that galaxy mergers normally reduce the stellar angular momentum of ETGs.

Different types of tidal features (e.g., shells, tails, and streams) store information about recent mergers with different properties. For example, shells can be generated by mergers with radial orbits, while streams can be produced by mergers with circular orbits \citep{Quinn1984,Dupraz1986,Johnston2008,Hendel2015,Pop2018,Karademir2019}. Arm- and loop-shaped tidal tails may be formed by the dynamically cold material in the disks of merging galaxies \citep{Schombert1990,Feldmann2008,Duc2015}. Therefore, splitting tidal features into different types and exploring how these types are associated with the properties of galaxies can provide further insight into processes of galaxy mergers and evolution.

In this study, we extend the study of \citet{Yoon2022} by increasing the number of ETGs to over a thousand using the Dark Energy Spectroscopic Instrument (DESI) Legacy Imaging Survey \citep{Dey2019}, which has a comparable surface brightness limit with that of the Stripe 82 coadded images but covers a far larger survey area. Using this substantially larger sample, we focus on examining how the stellar angular momentum of ETGs is different from each other, depending on the presence of the different types of tidal features. By doing so, we broaden our understanding of the impact of galaxy mergers on the stellar kinematics of ETGs.

 In this study, $H_0=70$ km s$^{-1}$ Mpc$^{-1}$, $\Omega_{\Lambda}=0.7$, and $\Omega_\mathrm{m}=0.3$ are used as the cosmological parameters. 
\\

\section{Sample and Analysis}\label{sec:sample}

\subsection{SDSS-IV MaNGA}\label{sec:manga}
The MaNGA IFU spectroscopic survey is the fourth generation of the SDSS project \citep{Blanton2017}. This project collected observational data using the Astrophysical Research Consortium (ARC) 2.5m telescope. The MaNGA project employed 17 hexagonal shape fiber-bundled IFUs with sizes ranging from $12\arcsec$ -- $32\arcsec$, depending on the number of fibers. These 17 IFUs are distributed across the $3\degr$ field of view of the telescope focal plane. The spectrograph utilized in the MaNGA survey is identical to the one used in the Baryon Oscillation Spectroscopic Survey \citep{Smee2013}. This spectrograph covers a wavelength range of 3600--10300\AA\,, offering a midrange spectral resolution of $R\sim2000$. The target selection criteria of the MaNGA survey is based on $i$-band absolute magnitude and redshift (and near-ultraviolet$- i$ color for a small portion of targets). This selection process results in $\sim10,000$ galaxies, evenly distributed across the color-magnitude space, with a uniform spectroscopic coverage up to 1.5 or 2.5 half-light radius along the major axis ($R_e$). Further details regarding the selection of target galaxies can be found in \citet{Wake2017}.

Here, we utilize the MaNGA data of the final release version (Data Release 17). Approximately $1\%$ of the MaNGA data were obtained from repeated observations of the same galaxy. In these cases, we select one observation from the duplicates, prioritizing the data with the larger IFU size. If the IFU sizes of the duplicate observations are identical, we choose the data with the highest blue channel signal-to-noise ratio (S/N).
\\

\subsection{Deconvolution of IFU Data}\label{sec:decon}
Observations from ground-based telescopes are inevitably influenced by instrument-induced aberrations and atmospheric conditions, causing the seeing effect. The effect of seeing is particularly pronounced in data obtained through fiber-based IFUs, owing to the large physical gaps between the sampling elements (IFU fibers). Reducing this effect enables us to derive more reliable spatially resolved kinematics information, especially in the central regions of galaxies. In order to alleviate the impact of seeing, we apply the Lucy--Richardson (LR) deconvolution algorithm \citep{Richardson1972,Lucy1974} to the MaNGA IFU data as in \citet{Chung2021} (the same algorithm was also used in \citealt{Yoon2021} and \citealt{Yoon2022}). The LR deconvolution algorithm is an iterative process designed to restore an original image that has undergone convolution by a point spread function (PSF). This algorithm is characterized by its minimal parameter requirements, making it well suited for the deconvolution of very large datasets. Comprehensive information regarding the algorithm and its application to MaNGA data cubes can be found in \citet{Chung2021}. Thus, in this paper, we provide a concise description of the application of the LR algorithm to the IFU data.

The LR algorithm can be expressed using a simple equation,
\begin{equation}
u^{n+1}=u^n\cdot \Bigg[\bigg(\frac{d}{u^n\otimes p}\bigg)\otimes p\Bigg],
\label{eq:deconv}
\end{equation}
where $u^n$ represents the $n\mathrm{th}$ estimate of the maximum likelihood solution, while $d$ stands for the original PSF-convolved image (hence $u^0=d$). The parameter $p$ denotes a two-dimensional (2D) PSF, and the symbol $\otimes$ denotes 2D convolution. 

The LR algorithm is applied to the MaNGA cube data by deconvolving the 2D image slice at each wavelength bin individually. The MaNGA data cubes provide PSF full width half-maximum (FWHM) values for the $g$, $r$, $i$, and $z$ bands. We conduct a linear fitting on these FWHM values at the respective wavelengths of the four bands. The FWHM of the PSF at each wavelength bin in data cubes is derived from this linear model through interpolation.\footnote{The difference between the FWHM values of the four bands in the MaNGA data cubes and those obtained from the fitted linear function is extremely small, with an average absolute difference of only $0.007\arcsec$\citep{Chung2021}. This is attributed to the fact that the FWHM value in MaNGA data shows only a weak dependence on wavelength \citep{Law2016,Chung2021}. Consequently, the FWHM value derived from the linear function provides an accurate approximation at each wavelength bin.} In the deconvolution process for a specific wavelength, the 2D Gaussian function with the interpolated FWHM value is applied.\footnote{The PSFs in MaNGA data can be well modeled by a single 2D Gaussian function, with the FWHM showing a variation of less than $10\%$ across a given IFU \citep{Law2015,Law2016}.} The number of iterations ($N_\mathrm{iter}$) in the LR algorithm is set to 20, which is determined to be optimal through the tests in \citet{Chung2021}. Going beyond $N_\mathrm{iter}=20$ does not lead to a significant enhancement in deconvolution quality; instead, it introduces additional artifacts in the image with amplified noise. 

The test for the applications of the LR algorithm to simulated IFU data in \citet{Chung2021} demonstrates that the deconvolution process enables us to effectively recover the true stellar kinematics of galaxies. For example, the test indicates that the luminosity-weighted stellar angular momentum can be restored with an underestimation of less than $\sim5\%$ in the majority of cases where the deconvolution process is used. By contrast, without applying deconvolution, the level of underestimation can reach $20\%$--$30\%$.
\\

\begin{figure*}
\hspace*{-0.45cm}  
\includegraphics[width=1.046\linewidth]{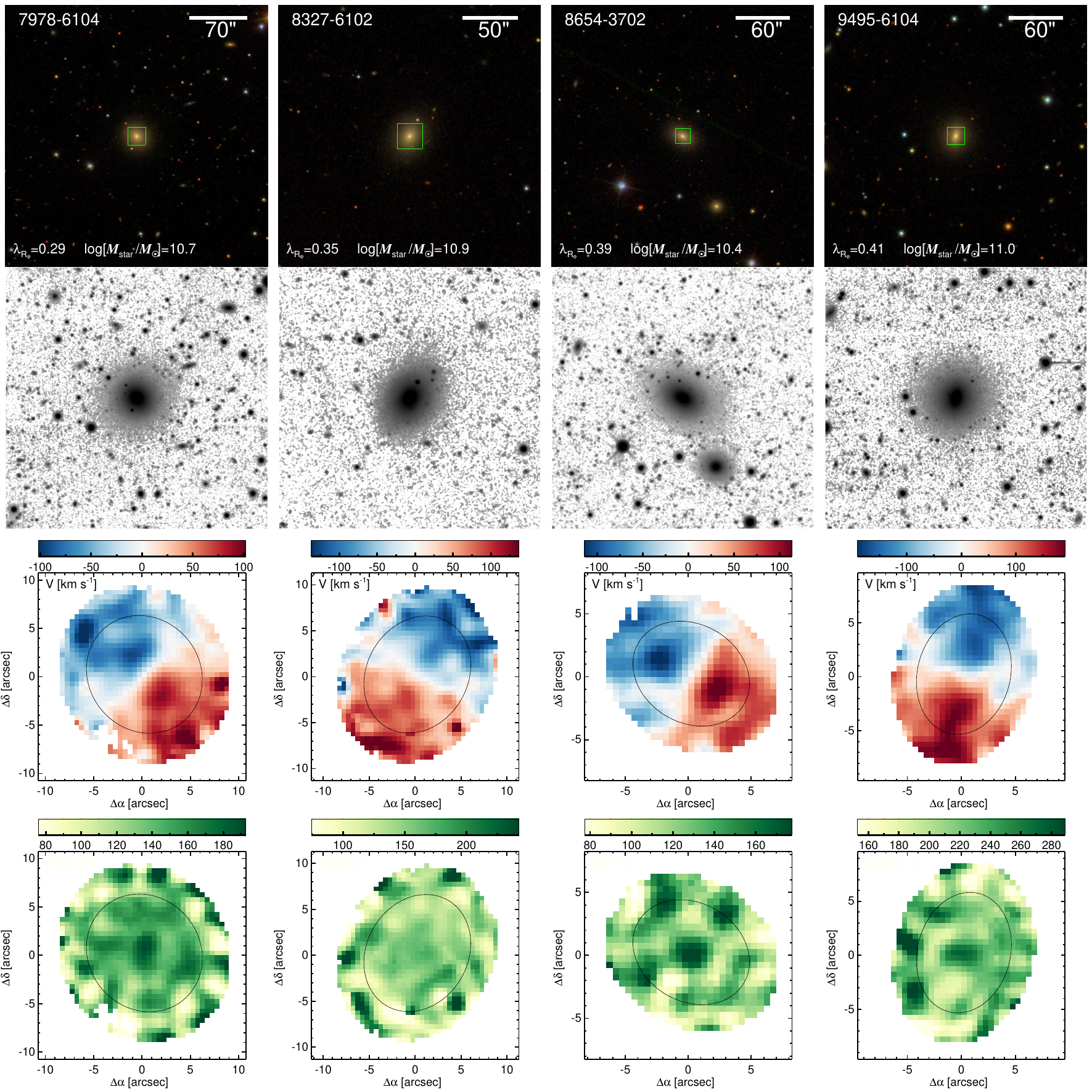}  
\centering
\caption{Examples of normal ETGs without tidal features. First row: color images from SDSS. The plate ID (e.g., 7978) and IFU design ID (e.g., 6104) are shown in the color image. The horizontal bar in the color image denotes the angular scale of the image. The green square denotes the window size of the 2D line-of-sight velocity and velocity dispersion maps shown in the third and fourth rows. The $\lambda_{R_e}$ and $\log(M_\mathrm{star}/M_{\odot})$ of each galaxy are shown in the bottom of each color image. Second row: $r$-band deep images of the DESI Legacy Survey. The angular scale of the deep image is the same as that of the color image. Third row: 2D line-of-sight velocity maps within $1.5R_e$. Fourth row: 2D line-of-sight velocity dispersion maps within $1.5R_e$. The black ellipses in the third and fourth rows indicate the areas within which $\lambda_{R_e}$ values are calculated. The color bars positioned above the panels in the third and fourth rows represent the color-coded velocity and velocity dispersion scales, respectively. The symbols $\Delta\alpha$ and $\Delta\delta$ denote relative R.A. and decl., respectively. 
\label{fig:ex_1}}
\end{figure*}

\begin{figure*}
\hspace*{-0.45cm} 
\includegraphics[width=1.046\linewidth]{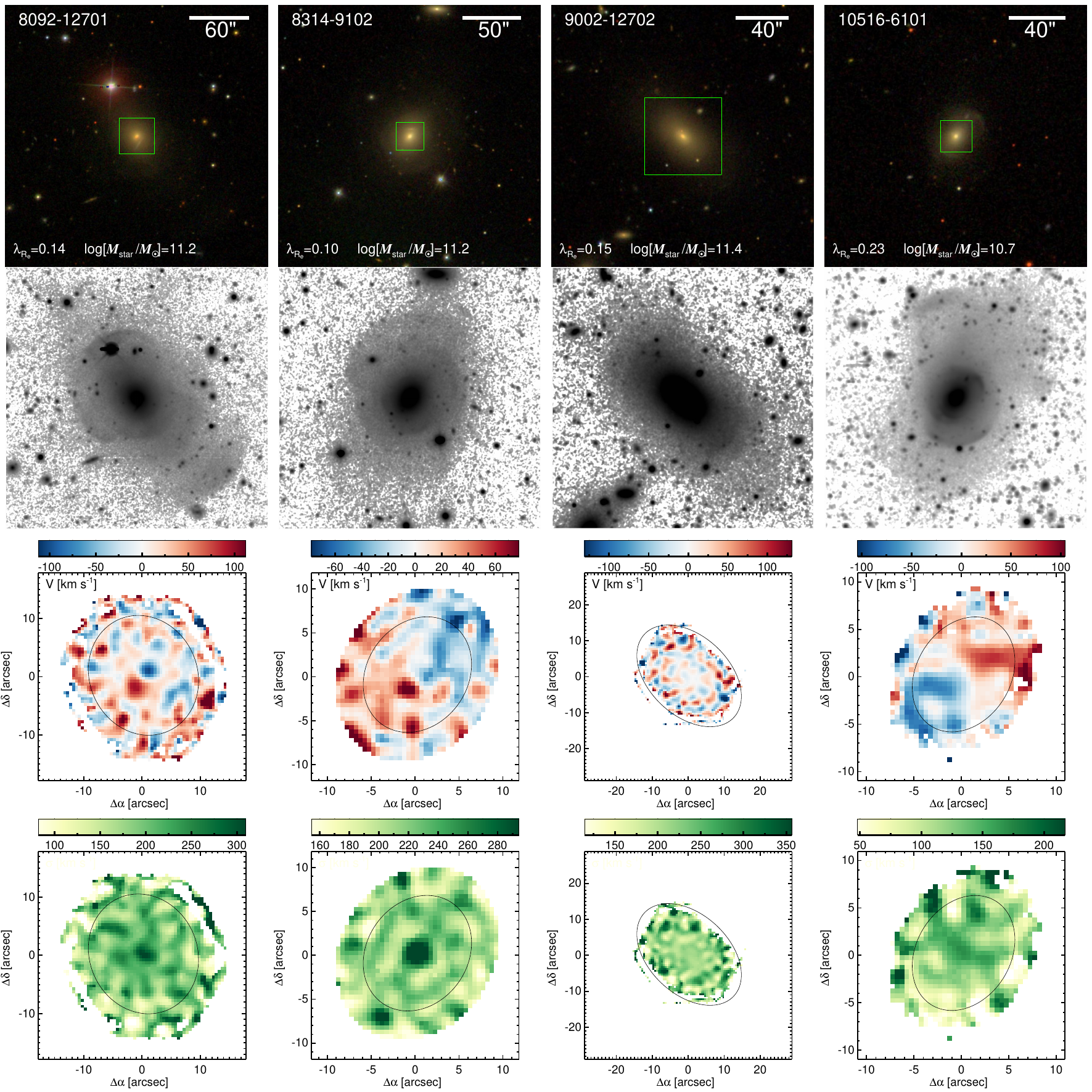}
\centering
\caption{Examples of ETGs with shell-type tidal features. The figure descriptions are the same as in Figure \ref{fig:ex_1}.
\label{fig:ex_2}}
\end{figure*} 

\begin{figure*}
\hspace*{-0.45cm} 
\includegraphics[width=1.046\linewidth]{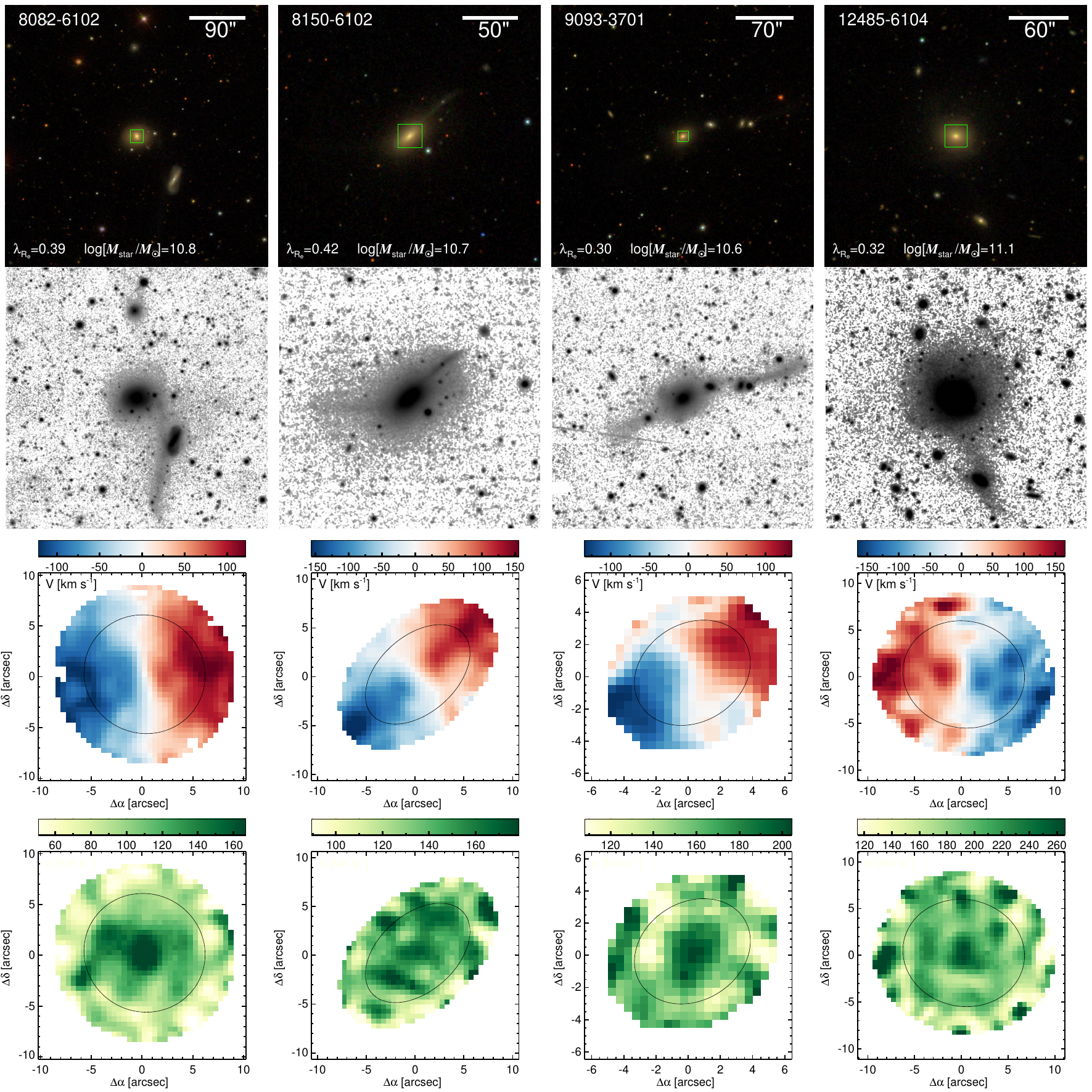}  
\centering
\caption{Examples of ETGs with stream-type tidal features. The figure descriptions are the same as in Figure \ref{fig:ex_1}.
\label{fig:ex_3}}
\end{figure*} 

\begin{figure*}
\hspace*{-0.45cm} 
\includegraphics[width=1.046\linewidth]{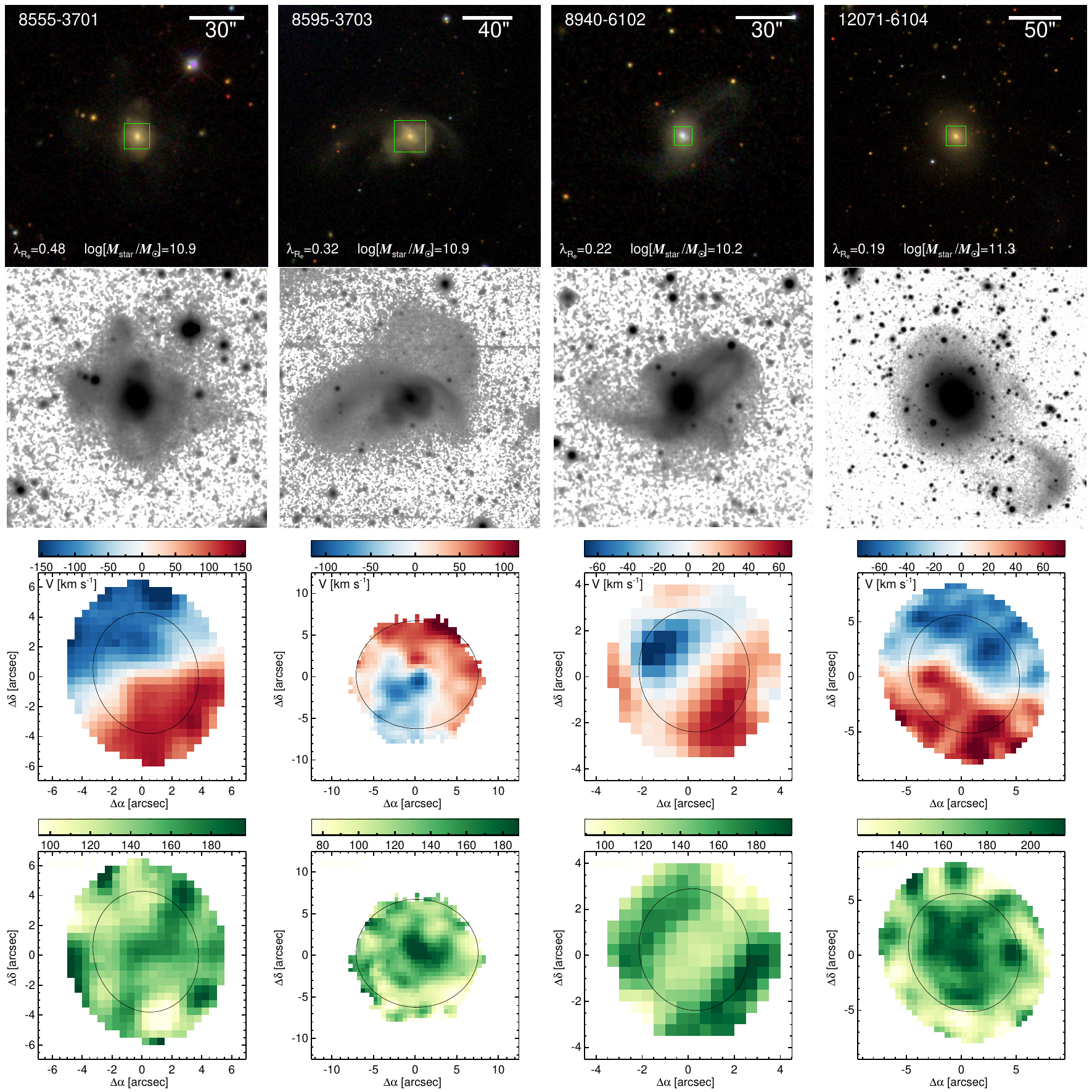}  
\centering
\caption{Examples of ETGs with tidal features. These ETGs have both tails and streams. The figure descriptions are the same as in Figure \ref{fig:ex_1}.
\label{fig:ex_4}}
\end{figure*} 

\begin{figure*}
\hspace*{-0.45cm} 
\includegraphics[width=1.046\linewidth]{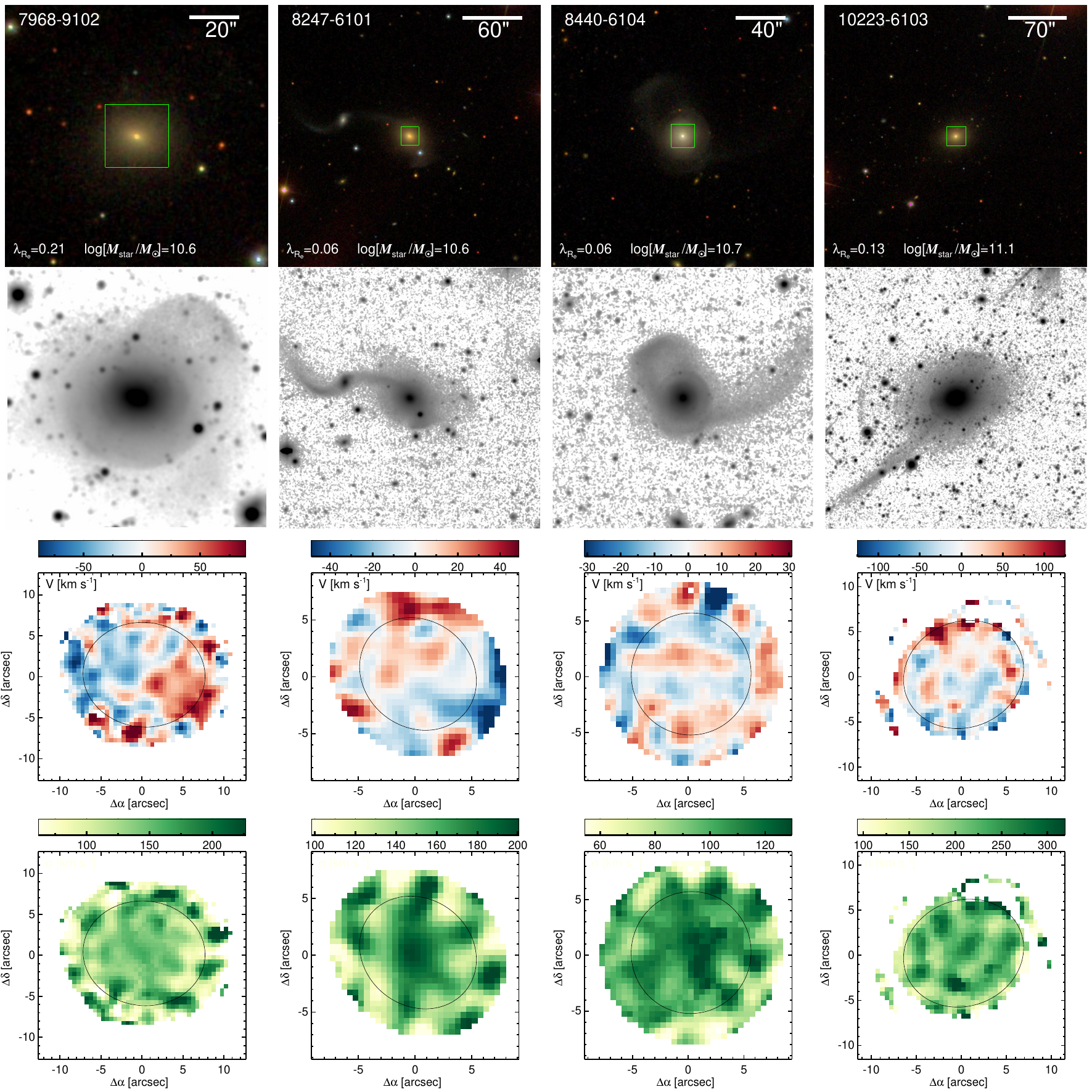}  
\centering
\caption{Examples of ETGs with tidal features. These ETGs have both shells and streams (galaxy 8440-6104 also displays a tail-type tidal feature). The figure descriptions are the same as in Figure \ref{fig:ex_1}.
\label{fig:ex_5}}
\end{figure*}

\subsection{Extracting Stellar Kinematics}\label{sec:extract}
We use the method outlined in \citet{Yoon2021} to derive stellar kinematics from the MaNGA data, such as line-of-sight velocities and velocity dispersions. The code used here for extracting stellar kinematics is the Penalized Pixel-Fitting (pPXF) method, which is designed to use the maximum penalized likelihood formalism for full spectrum fitting on galaxy spectra. We utilize MILES single stellar population models \citep{Sanchez2006,Vazdekis2010,Falcon2011} with the Padova+00 isochrone \citep{Girardi2000} and the initial mass function of \citet{Chabrier2003} for input model templates in pPXF. The model templates encompass six metallicity values spanning $-1.71\leq\mathrm{[M/H]}\leq0.22$\footnote{As mentioned in \citet{Yoon2022}, the use of alternative MILES models derived from the BaSTI isochrone \citep{Hidalgo2018}, which covers up to a higher metallicity of $[M/H]=0.4$, yields similar kinematics for the most massive ETGs with $\log(M_\mathrm{star}/M_\odot)>11.2$, whose metallicities are expected to be high.} and 10 age values ranging from 0.07--12.59 Gyr, totaling 60 model templates. Following the approach of \citet{Belfiore2019}, we incorporate an eighth-order additive Legendre polynomial into the fit to improve the quality of the derived stellar kinematics (see \citealt{Emsellem2004}).

As in the method of \citet{Cappellari2017}, the model templates are convolved with a Gaussian function to align with the resolution of the MaNGA spectra. Additionally, the spectra are shifted to the rest frame before extracting stellar kinematics. The fitting process is carried out after masking the pixels around known emission lines and bad pixels (such as those with low coverage depth, dead fibers, or contamination from foreground stars, etc.) flagged in the MaNGA data reduction pipeline \citep{Law2016}. The fitting range is confined to the wavelength range of 3700--7400\AA\, to match the wavelength coverage of the model templates (3540--7410\AA). In the third and fourth rows of Figures \ref{fig:ex_1}--\ref{fig:ex_5}, we present examples of the 2D line-of-sight stellar velocity and velocity dispersion maps for our ETG sample.

We calculate the dimensionless spin parameter luminosity-weighted specific stellar angular momentum, $\lambda_R$ following the method in \citet{Emsellem2007},
\begin{equation}
\lambda_R\equiv\frac{\langle R \, \vert V \vert \rangle}{\langle R \, \sqrt{V^2+\sigma^2} \rangle }=\frac{\sum^{N}_{i=1}F_i\,R_i\, \vert V_i \vert}{\sum^{N}_{i=1}F_i\,R_i \, \sqrt{V_i^2+\sigma_i^2}},
\label{eq:lambda}
\end{equation}
in which $F_i$ represents the flux of the $i$th bin, while $R_i$ denotes the circular radial distance from the center to the $i$th bin. Here, we use the $F_i$ of $r$-band images, which is deconvolved in the same manner as with IFU data. As for $V_i$ and $\sigma_i$,  these refer to the line-of-sight velocity and velocity dispersion of the $i$th bin, respectively. The summation is performed over $N$ pixels located within the photometric ellipse. The spin parameter $\lambda_R$ is scaled by $\sqrt{V_i^2+\sigma_i^2}$, which serves as a proxy for mass. The parameter $\lambda_R$ approaches unity in the cases where the system is rotation dominated, and converges to zero when the system is predominantly pressure supported (dominated by random motions of stars). For the computation of $\lambda_R$, we consider only spaxels with a median S/N $\ge5$. We do not use the spaxels with $\sigma<40$ km s$^{-1}$ in the calculation, as such low $\sigma$ values can be unreliable due to the instrumental resolution limit \citep{Penny2016,Lee2018}.\footnote{In \citet{Yoon2022}, we tested the effect of excluding spaxels with low $\sigma$ on the derived $\lambda_R$ values. This was done by setting $\sigma$ values to $40$ km s$^{-1}$ for spaxels with $\sigma<40$ km s$^{-1}$ in the calculation of $\lambda_R$. Through the test, we found that the prescription excluding spaxels $\sigma<40$ km s$^{-1}$ essentially does not change the calculated $\lambda_R$.} We further remove spurious spaxels with $\vert V \vert\ge500$ km s$^{-1}$ in the calculation of $\lambda_R$. 

The parameter $\lambda_R$ exhibits little variation over a broad range of viewing angles, except when the angle is nearly face-on \citep{Emsellem2007,Jesseit2009,Bois2011}. This stability arises from the concurrent decrease of $\langle V \rangle$ and $\langle \sigma \rangle$ as the inclination decreases, resulting in only minor variations in their ratio \citep{Jesseit2009}. Consequently, $\lambda_R$  serves as a robust observational indicator for the intrinsic angular momentum in the majority of galaxies.

For this reason, $\lambda_R$ has been widely used in many previous studies examining the stellar angular momentum of galaxies \citep{Jesseit2009,Emsellem2011,Fogarty2015,Cappellari2016,Oh2016,Choi2017,Graham2018}. These works have used $\lambda_R$ within $R_e$ (hereafter, $\lambda_{R_e}$) for statistical analyses of galaxy stellar angular momentum, and we also use $\lambda_{R_e}$ throughout this study. Here, the parameters $R_e$ and ellipticities ($\varepsilon$) of galaxies are derived using the elliptical Petrosian flux in the $r$ band.\footnote{The parameter $\varepsilon$ is calculated at $90\%$ light radius.} The stellar masses ($M_\mathrm{star}$) of the galaxies used in this study are likewise computed from the elliptical Petrosian fluxes. These parameters are sourced from the NASA-Sloan Atlas catalog, which is the base catalog for selecting target galaxies in the MaNGA project \citep{Wake2017}.
\\

\subsection{ETG Sample}\label{sec:sam}

We limit our sample to galaxies within the redshift range of $z<0.055$.  We exclude galaxies at higher redshifts, due to their small angular sizes and the cosmological surface brightness dimming effects (see Equation 6 in \citealt{YP2020}), which render the detection of tidal features challenging. As in \citet{Yoon2022}, we use galaxies with $M_\mathrm{star}\ge10^{9.65}\,M_{\odot}$. Note that tidal features are rarely detected at $M_\mathrm{star}<10^{9.65}\,M_{\odot}$ \citep{Yoon2022}. The number of MaNGA galaxies after implementing the redshift and stellar mass criteria is 5745.

For the selection of ETGs among MaNGA galaxies, we use a value-added catalog (MaNGA Visual Morphologies from SDSS and DESI images; \citealt{Vazquez2022}), which contains morphology classification information based on visual inspection of SDSS and DESI images, for all galaxies in MaNGA Data Release 17. The morphology classifications of this catalog are in good agreement with the visual classifications of \citet{Nair2010} and the machine-learning-based results of \citet{Dominguez2022}, showing median scatters in T type of 1.2 and 1.48, respectively \citep{Vazquez2022}. We double-check 1838 ETGs (T type$\le0$ in the catalog) through visual inspection of SDSS and DESI color images and exclude 26 galaxies, as they have morphologies more similar to late-type galaxies.

In order to use reliable $\lambda_{R_e}$ values in our analysis, we do not include IFU data where over $25\%$ of spaxels within $\lambda_{R_e}$ are excluded, due to the conditions described in Section \ref{sec:extract}, such as median S/N $<5$, $\sigma<40$ km s$^{-1}$, $\vert V \vert\ge500$ km s$^{-1}$, or contamination from foreground stars, etc. We additionally rule out IFU data where the total number of spaxels within $R_e$ is fewer than 45. The number of remaining ETGs is 1312, after excluding such IFU data. Finally, 68 ETGs are excluded due to poor image quality resulting from their proximity to bright sources (Section \ref{sec:tf}). As a result, the final sample consists of 1244 ETGs.
\\

\subsection{Detection and Classification of Tidal Features}\label{sec:tf} 
For the detection of tidal features, we use images of the DESI Legacy Survey Data Release 10 \citep{Dey2019}. The DESI Legacy Survey is a combination of three wide-area surveys, which are the Dark Energy Camera Legacy Survey, the Beijing-Arizona Sky Survey, and the Mayall $z$-band Legacy Survey. These surveys cover a total area of $\sim14,000\,\mathrm{deg}^2$. The median surface brightness limit ($1\sigma$ of the background noise over a $1\arcsec\times1\arcsec$ region) of $g$- and $r$-band DESI images is $\sim27$ mag arcsec$^{-2}$. This limit is similar to that of deep coadded $r$-band images of the Stripe 82 region of SDSS \citep{YL2020,Yoon2022,Yoon2023}, which have been widely used to identify low-surface brightness tidal features around galaxies (e.g., \citealt{Kaviraj2010,Schawinski2010,Hong2015}).

We perform a visual inspection of the $g$-band and $r$-band images, and composite color images of the $g$, $r$, and $z$ bands to identify tidal features. If necessary, especially when examining the inner regions of galaxies, we also inspect the residual images of the $g$ and $r$ bands, in which the 2D light models of galaxies\footnote{A de Vaucouleurs model, an exponential disk model, and a composite of the two models are used.} are subtracted from the original images. During the visual inspection of the images, we fine-tune the scale of the pixel values and enhance signals by smoothing images using Gaussian kernels of various sizes for the better identification of diffuse and faint tidal features. In this process, we find 68 galaxies that are located too close to bright stars and large galaxies. We exclude them, as mentioned in Section \ref{sec:sam} since it is difficult to identify tidal features in the images with the high background levels caused by nearby very bright sources. 

In \citet{YL2020} and \citet{Yoon2022}, our classifications of tidal features in the Stripe 82 region are compared with those made by \citet{Kaviraj2010}, who classified ETGs with $M_r\footnote{$M_r$: $r$-band absolute magnitude}<-20.5$ at $z<0.05$ in the Stripe 82 region into normal (relaxed) ETGs and ETGs with tidal features. The comparison reveals that over $\sim90\%$ of the classifications are in agreement with each other (see \citealt{YL2020} and \citealt{Yoon2022} for more details). 

We find that 254 out of 1244 ETGs have tidal features ($20.4\%$). This fraction is slightly lower than (but agrees with) the values of $22.8\%\pm3.0\%$ \citep{Yoon2023} and $24.6\%\pm3.3\%$ \citep{Yoon2022}, which were determined based on MaNGA ETGs with comparable stellar masses, utilizing deep coadded images of the Stripe 82 region. 

We classify the detected tidal features into three types, which are tails, streams, and shells. These are commonly used categories in previous studies \citep{Duc2015,Mancillas2019,Bilek2020,Bilek2023,Sola2022}. Tidal tails are thick, radially elongated structures that are visibly connected to the host galaxy. These elongated stellar features, potentially formed during major mergers, are morphologically similar to streams but possess a higher thickness, sometimes reaching the size of the host galaxy itself. However, there is no obvious visual difference between tails and streams in some cases \citep{Bilek2020,Sola2022}. Tidal streams are thin, elongated structures that usually resemble narrow filaments, likely tracing minor mergers. In some cases, they are connected to smaller companion galaxies. Shells are concentric, arc-shaped features with sharp edges. The alignment of arc features can either follow a common axis or appear to be randomly spread around the host galaxy. As shells extend to larger radii, they become more diffused. The exact conditions for their formation remain a topic of debate, with radial mergers generally being favored \citep{Quinn1984,Dupraz1986,Johnston2008,Hendel2015,Pop2018,Karademir2019}. If a galaxy has more than two types of tidal features, we assign the galaxy to all corresponding groups. For example, ETGs both with shell and stream features are classified as ETGs with shell-type tidal features, and at the same time, as ETGs with streams.

We evaluate the robustness of the separation of tidal features by combining independent categorizations provided by two authors (W.B. and K.C.) and adjusting the initial classifications (established by Y.Y. with assistance from J.K. and H.C.) based on the majority rule. We find that $93.7\%$ (238/254) of the tidal feature separations remain unchanged from the initial classifications when decisions are made using the majority rule among the three sources. Only 14 out of the 254 cases ($5.5\%$) are modified from the initial classifications.\footnote{One ETG with a shell feature is reclassified as an ETG with a stream feature. Nine ETGs with streams are changed to seven ETGs with tails and two ETGs with tails+streams. One ETG with a shell+stream feature is modified to an ETG with a shell+tail feature. Three ETGs with tails+streams are reclassified as three ETGs with streams.} For the remaining two cases, where all classifications from the three sources differ from each other, we follow the initial categorizations. The outcome of this process indicates that the categorization of tidal features into three types is quite robust, even though it relies on visual inspections that can be subjective. 

The examples of shallow SDSS color images and deep DESI images for ETGs without tidal features are shown in the first and second rows of Figure \ref{fig:ex_1}.  The examples for ETGs with tidal features are displayed in Figures \ref{fig:ex_2}--\ref{fig:ex_5} (Figure \ref{fig:ex_2}: ETGs with shells; Figure \ref{fig:ex_3}: ETGs with streams; Figure \ref{fig:ex_4}: ETGs with tails + streams; Figure \ref{fig:ex_5}: ETGs with shells + streams/tails).

The number of ETGs with the tidal features of tails, streams, and shells are 59, 177, and 72, respectively. The numbers of ETGs that have a purely single type of tidal features are 30, 128, and 45, for those with tails, streams, and shells, respectively. There are 27 ETGs that have both tails and streams, while 25 ETGs have both streams and shells. Five ETGs have tail- and shell-type tidal features. Three ETGs exhibit all three types of tidal features simultaneously.

Image depth can introduce potential biases in the detection of tidal features, thereby influencing our results. For example, the simulation of \citet{Mancillas2019} suggests that the detection of streams can be largely dependent on the surface brightness limit of images. In order to roughly figure out the impact of image depth on our results, we divide the 254 ETGs with tidal features into two groups: one group consists of ETGs with prominent tidal features that are clearly visible without the need for smoothed or residual images, or are even visible in shallow images of SDSS (151 ETGs); the other group comprises the remaining ETGs with faint tidal features (103 ETGs). We find that the respective results obtained based on these two groups are consistent with the main findings derived from the full sample. For example, the median $\lambda_{R_e}$ for ETGs with prominent tidal features is $0.20\pm0.01$. The median $\lambda_{R_e}$ values for ETGs with prominent shells, tails, and streams are $0.13\pm0.02$, $0.21\pm0.03$, and $0.22\pm0.02$, respectively. The fraction of slow rotators among ETGs with prominent tidal features is $0.27\pm0.04$, while the slow rotator fractions for those with prominent shells, tails, and streams are $0.47\pm0.07$, $0.20\pm0.06$, and $0.20\pm0.04$, respectively. These quantities, including those from faint tidal features, are consistent within the margins of error with those derived from the full sample specified in Section \ref{sec:results}. This implies that image depth is not a critical factor affecting our results within the depth range we can probe with DESI images. However, the use of much deeper images from future large-area surveys will enable us to more fully understand the influence of image depth.

Another factor that can produce potential biases is the projection angle. According to \citet{Mancillas2019}, the detection of streams and tails does not depend on projection angles, while the detection of shells can be more substantially affected by the projection. However, if only $\sim20\%$ of shells are missed when restricted to a single projection, as found in the simulations of \citet{Pop2018}, such dependency in shell detections is unlikely to greatly impact our main results. Moreover, if the detection of prominent tidal features with high surface brightness is less affected by projection angles, as shown by \citet{Martin2022}, the fact that the respective results derived from both the samples of prominent and faint tidal features do not show significant discrepancies from the overall findings from the full sample implies that biases due to projection angles may not be substantial enough to fundamentally alter our results.
\\

\begin{figure*}
\includegraphics[width=0.5\linewidth]{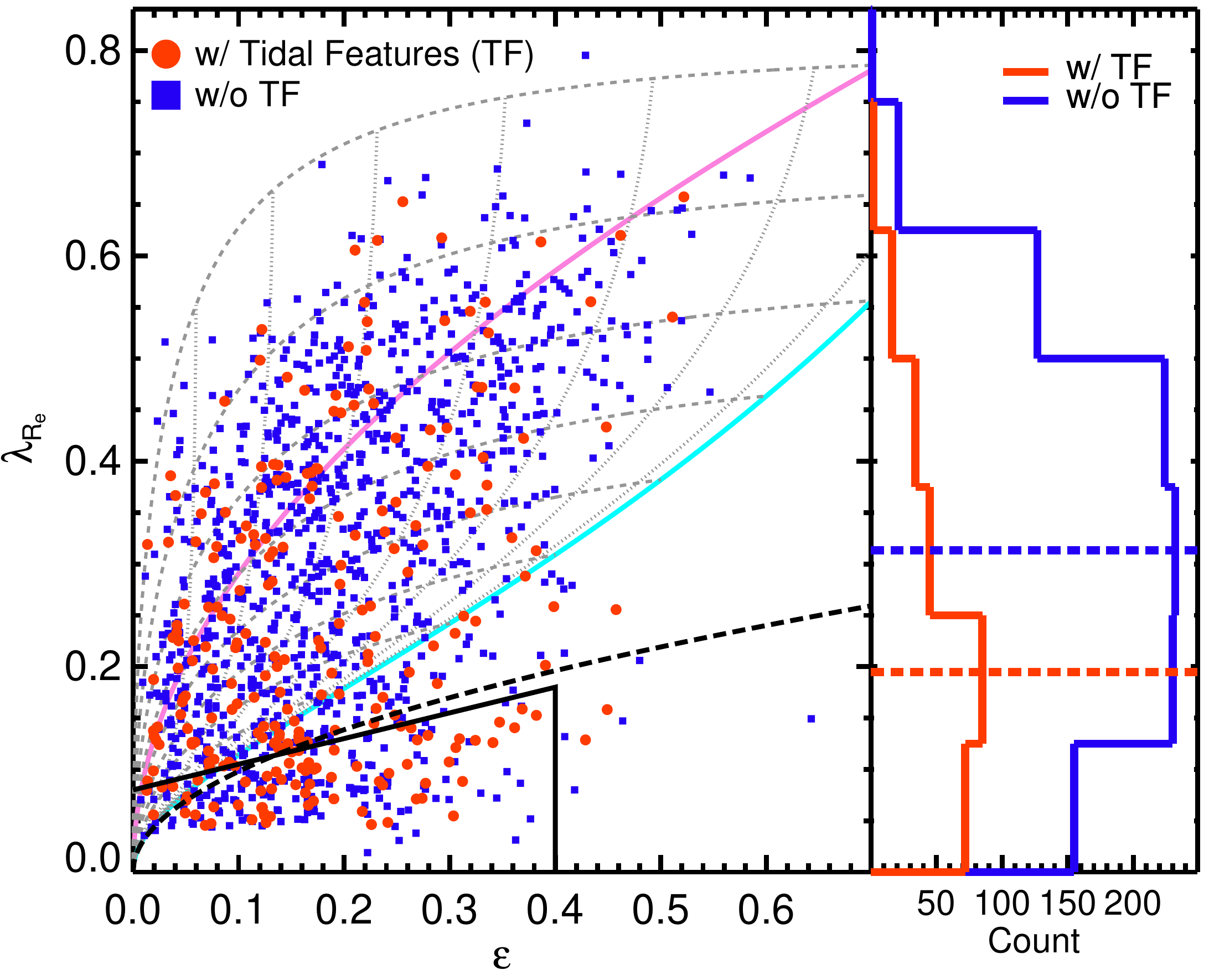}\includegraphics[width=0.5\linewidth]{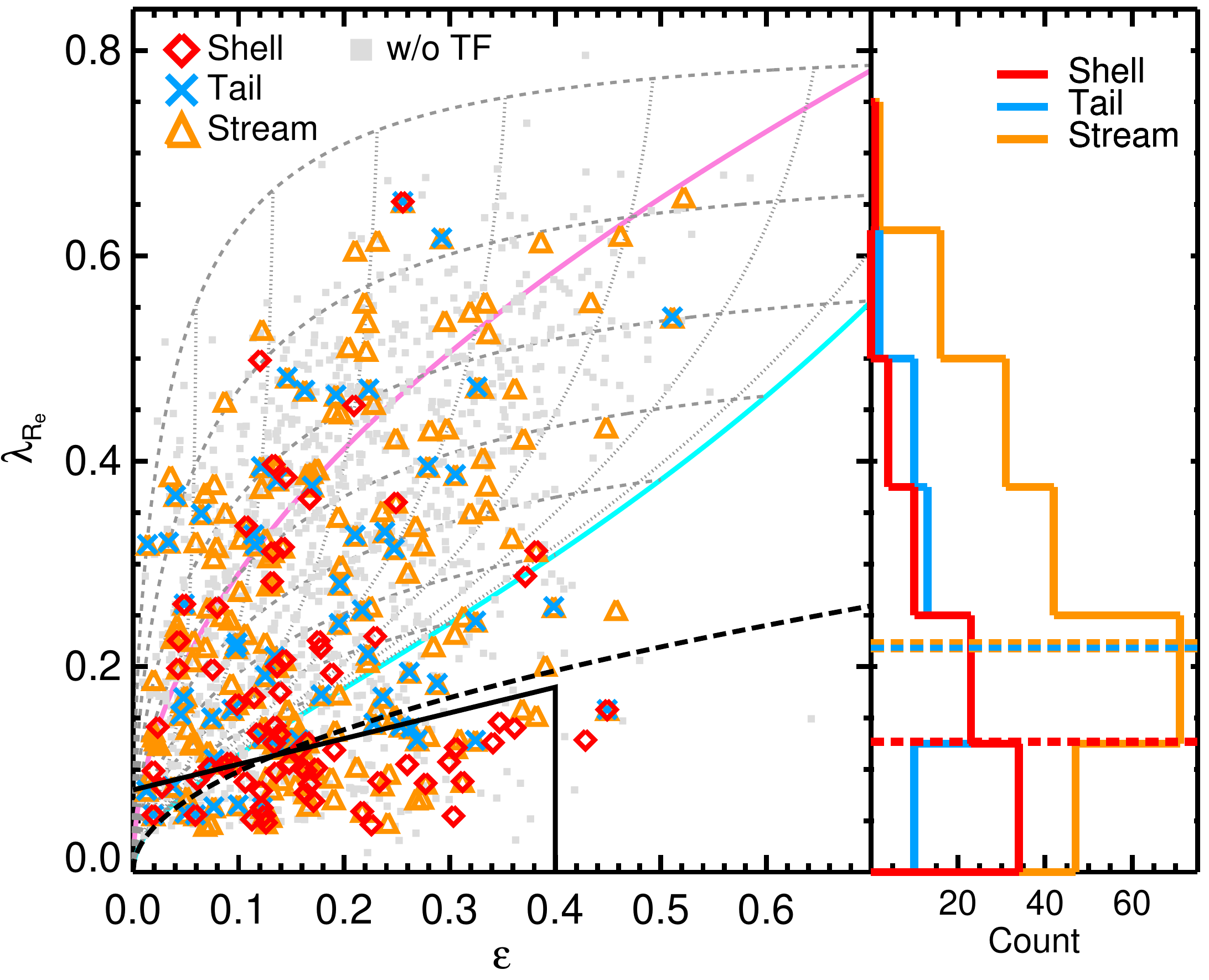}  
\centering
\caption{Distribution of ETGs in the $\lambda_{R_e}$ versus $\varepsilon$ plane. In the left panel, ETGs are divided into two categories based on the presence of tidal features. In the right panel, ETGs with tidal features are additionally subdivided into three categories based on the types of tidal features, as indicated in the legend. The right side of each panel exhibits a histogram of $\lambda_{R_e}$ for each ETG category, with the horizontal dashed lines indicating the median $\lambda_{R_e}$ for each respective ETG category. The solid black line and the black-dashed line represent the criteria for identifying slow rotators, as defined in \citet{Cappellari2016} and \citet{Emsellem2011}, respectively. The pink line on the $\lambda_{R_e}$ versus $\varepsilon$ plot represents $\lambda_{R_e}$ values for the edge-on isotropic rotator from \citet{Binney2005} with various intrinsic ellipticities \citep{Cappellari2016}. The light blue line denotes $\lambda_{R_e}$ values for edge-on galaxies following the anisotropy versus intrinsic flattening relation described in Equation 11 of \citet{Cappellari2016}. The gray-dotted lines indicate different versions of the light blue line at different inclination angles with a step size of $0.1\degr$. The gray-dashed line traces the path of galaxies with a constant intrinsic ellipticity as the inclination angle varies.
\label{fig:main}}
\end{figure*} 

\begin{figure*}
\includegraphics[width=0.5\linewidth]{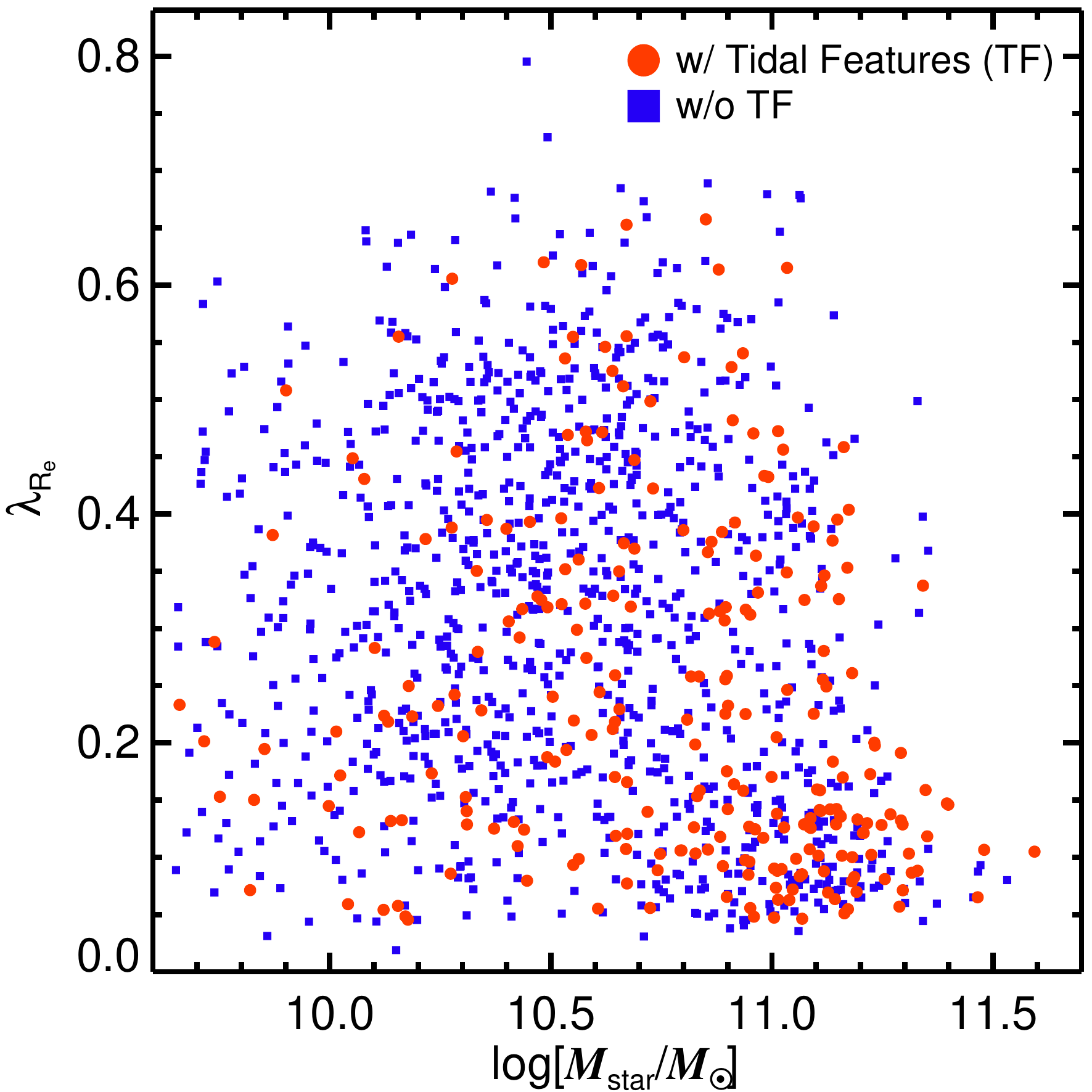}\includegraphics[width=0.5\linewidth]{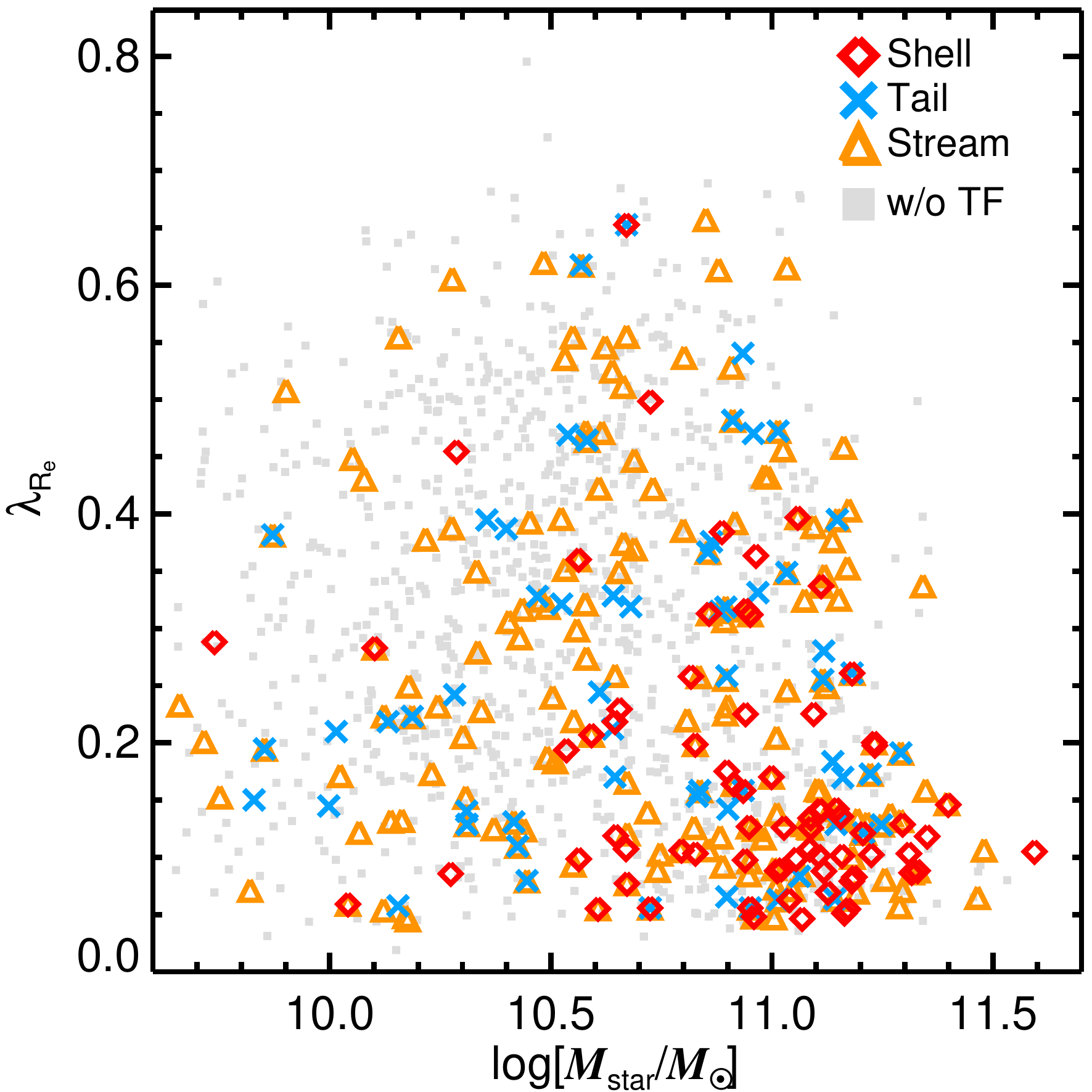}  
\centering
\caption{Distribution of ETGs in the $\lambda_{R_e}$ versus $\log M_\mathrm{star}$ plane. In the left panel, ETGs are separated into two categories based on the presence of tidal features. In the right panel, ETGs with tidal features are subdivided into three categories based on the types of tidal features, as indicated in the legend.
\label{fig:mass}}
\end{figure*}

\begin{figure}
\includegraphics[width=\linewidth]{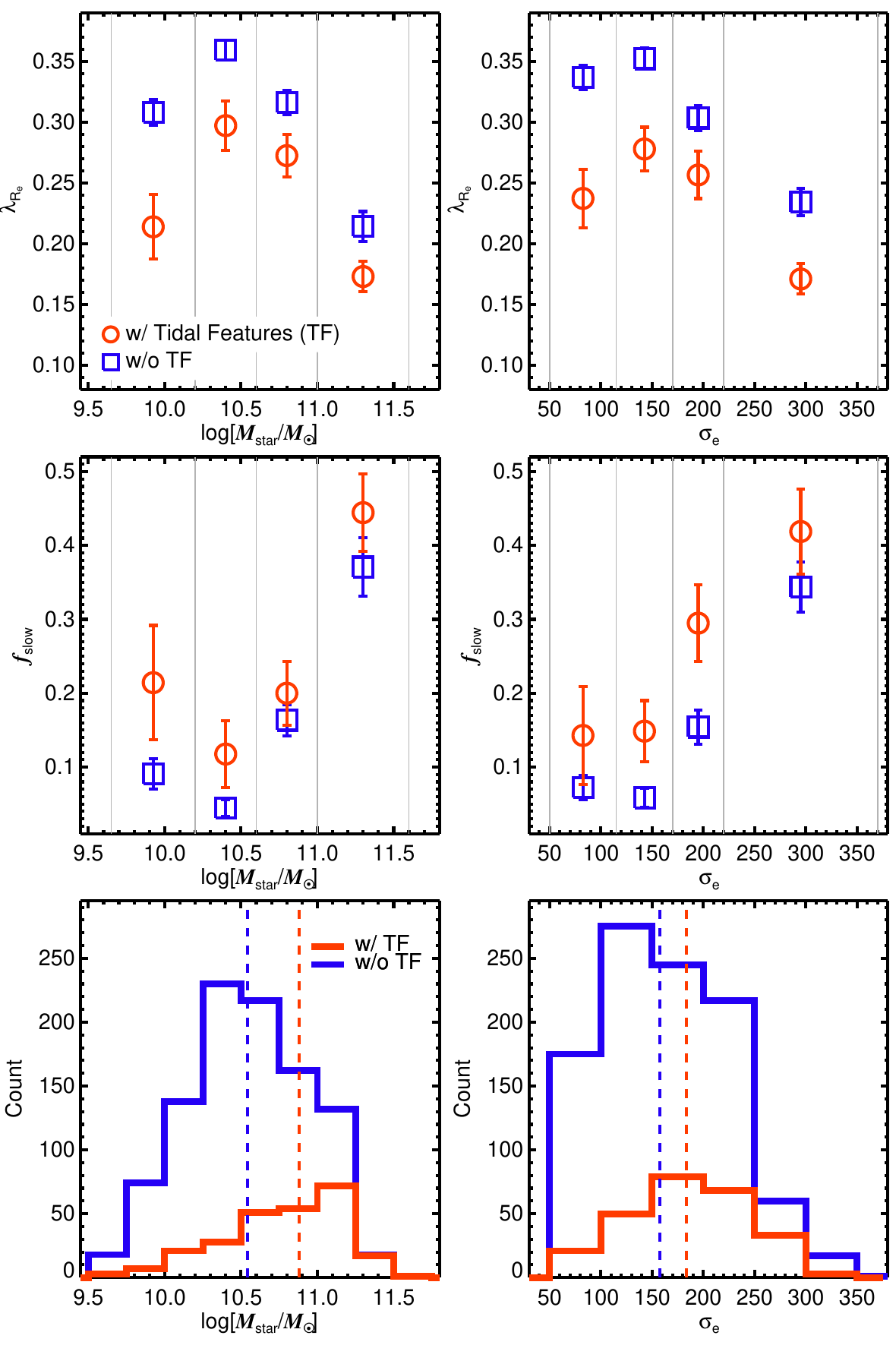}
\centering
\caption{Upper panels: mean $\lambda_{R_e}$ for ETGs with/without tidal features as a function of $\log M_\mathrm{star}$ and $\sigma_e$. The error bars indicate the standard error of the mean. Middle panels: the fraction of slow rotators ($f_\mathrm{slow}$) for ETGs with/without tidal features as a function of $\log M_\mathrm{star}$ and $\sigma_e$. Here, we use the criterion for defining slow rotators from \citet{Cappellari2016} (the solid black line in Figure \ref{fig:main}.) The error bars denote the standard error of the proportion. The gray vertical lines in the upper and middle panels indicate the boundaries of the bins used for calculating $f_\mathrm{slow}$ and mean $\lambda_{R_e}$. Lower panels: the histograms of $\log M_\mathrm{star}$ and $\sigma_e$ for ETGs with and without tidal features. The vertical dashed lines represent the median values of $\log M_\mathrm{star}$ or $\sigma_e$ for each ETG category.
\label{fig:bin_dls}}
\end{figure} 
\begin{figure}
\includegraphics[width=\linewidth]{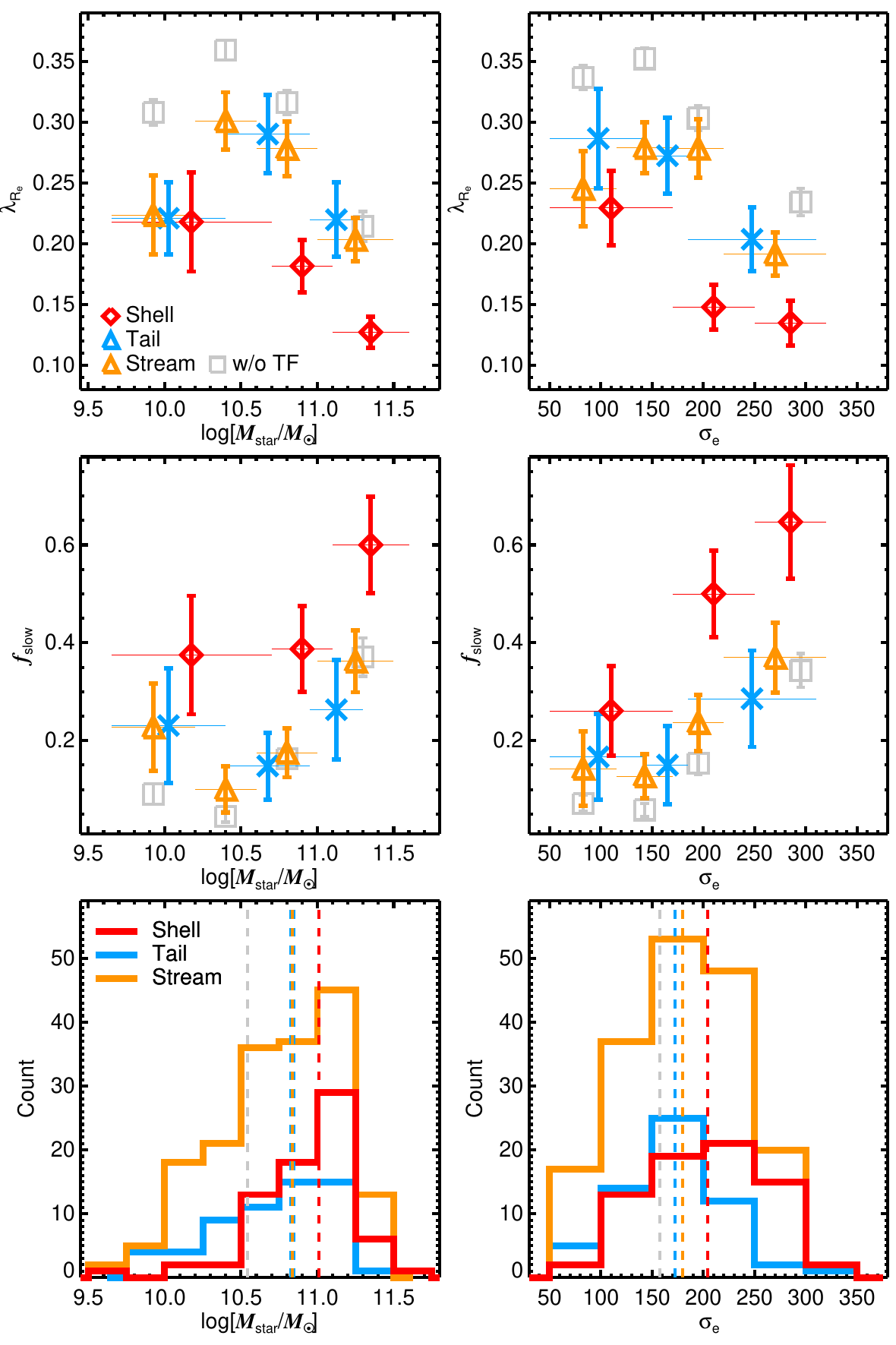}
\centering
\caption{This figure is identical to Figure \ref{fig:bin_dls}, except that ETGs with tidal features are subdivided into those with shells, tails, and streams. The horizontal bars in the upper and middle panels indicate the ranges of the bins used for computing $f_\mathrm{slow}$ and mean $\lambda_{R_e}$. The gray vertical dashed lines in the bottom panels indicate the median values for ETGs without tidal features. 
\label{fig:bin_class}}
\end{figure}

\section{Results}\label{sec:results}

In this section, we first present the results when ETGs are divided into those with and without tidal features. Following that, we demonstrate our novel findings when ETGs with tidal features are further divided into three categories based on the types of tidal features.

The left panel of Figure \ref{fig:main} displays the distribution of ETGs with/without tidal features in the $\lambda_{R_e}$ versus $\varepsilon$ plane and its projected histogram for $\lambda_{R_e}$. The figure exhibits that ETGs with tidal features have lower $\lambda_{R_e}$ than those without tidal features. The median $\lambda_{R_e}$ for ETGs with tidal features is $0.19\pm0.01$, whereas for ETGs without tidal features, it is $0.31\pm0.01$. We perform a Kolmogorov--Smirnov (KS) test for the distributions of $\lambda_{R_e}$ for the two ETG categories to assess the significance of the difference. The test yields that the probability ($0\le p \le1$) of the null hypothesis in which the two distributions originate from the same distribution is $p=6.0\times10^{-11}$, which indicates that the two $\lambda_{R_e}$ distributions are significantly different from each other.

The fraction of slow rotators ($f_\mathrm{slow}$) for the ETGs with/without tidal features, derived from the criterion of $\lambda_{R_e}<0.08+\varepsilon/4$ and $\varepsilon<0.4$ (under the solid black line in Figure \ref{fig:main}; \citealt{Cappellari2016}), is $f_\mathrm{slow}=0.27\pm0.03$ for ETGs with tidal features and $f_\mathrm{slow}=0.14\pm0.01$ for ETGs without tidal features.\footnote{When using another criterion of $\lambda_{R_e}<0.31\sqrt{\varepsilon}$ (under the black-dashed line in Figure \ref{fig:main}; \citealt{Emsellem2011}), the $f_\mathrm{slow}$ values are $0.26\pm0.03$ and $0.13\pm0.01$ for ETGs with and without tidal features, respectively.} This suggests that more than one-fourth of ETGs with tidal features are slow rotators, a fraction that is nearly twice as high as that of ETGs without tidal features.

The left panel of Figure \ref{fig:mass} presents the distribution of ETGs with/without tidal features in the $\lambda_{R_e}$ versus $\log M_\mathrm{star}$ plane. The upper panels of Figure \ref{fig:bin_dls} illustrate the mean $\lambda_{R_e}$ for ETGs with/without tidal features as a function of $\log M_\mathrm{star}$ and effective velocity dispersion within $R_e$ (hereafter $\sigma_e$). The parameter $\sigma_e$ is determined using Equation 3 in \citet{Graham2018}. These figures show that $\lambda_{R_e}$ of ETGs with tidal features are lower by $\sim0.06$ than that of ETGs without tidal features in all the $M_\mathrm{star}$ and $\sigma_e$ bins. KS tests on the distributions of $\lambda_{R_e}$ for the two ETG categories, corrected for the variation of $\lambda_{R_e}$ as a function of $M_\mathrm{star}$ and $\sigma_e$,\footnote{The correction is achieved by stacking the relative distributions of $\lambda_{R_e}$, with each distribution centered on (or normalized based on) the mean $\lambda_{R_e}$ of normal ETGs within each respective bin.} yields $p<1.1\times10^{-8}$, which demonstrates that the $\lambda_{R_e}$ values of the two ETG categories are still significantly different from each other even after accounting for the relation between $\lambda_{R_e}$ and mass. 

The fraction $f_\mathrm{slow}$ follows exactly the same trend as shown in the middle panels of Figure \ref{fig:bin_dls}, so that in all the $M_\mathrm{star}$ and $\sigma_e$ bins, $f_\mathrm{slow}$ of ETGs with tidal features is higher by $\sim0.08$ than that of ETGs without tidal features. We note that the mean $\lambda_{R_e}$ is the lowest; hence, $f_\mathrm{slow}$ is the highest in the most massive bin of $\log(M_\mathrm{star}/M_\odot)\ge11.0$, which is consistent with the results of previous studies \citep{Emsellem2007,Graham2018}. 

The histograms presented in the lower panels of Figure \ref{fig:bin_dls} indicate a higher occurrence of tidal features in more massive ETGs. The median $\log(M_\mathrm{star}/M_\odot)$ values are $10.88\pm0.03$ and $10.54\pm0.02$ for ETGs with and without tidal features, respectively. A KS test performed on the distributions of $\log(M_\mathrm{star}/M_\odot)$ for the two ETG categories provides $p=1.8\times10^{-15}$, indicating a highly significant difference between the two distributions. This result aligns with those reported in previous studies \citep{Bilek2020,Bilek2023,YL2020}.

Our results and trends regarding $\lambda_{R_e}$ when ETGs are separated into those with and without tidal features are almost identical to those presented in \citet{Yoon2022}. Therefore, by using a sample of ETGs more than six times larger than that used in \citet{Yoon2022}, we verify the principal findings of \citet{Yoon2022}. 

Now, we present the results when ETGs with tidal features are subdivided into three categories based on the types of tidal features, such as tails, streams, and shells. The right panel of Figure \ref{fig:main} shows the distribution of ETGs with shells, tails, and streams, as well as those without tidal features, in the $\lambda_{R_e}$ versus $\varepsilon$ plane. The histograms of the $\lambda_{R_e}$ distribution for each ETG category are shown on the right side of the panel. Figure \ref{fig:main} demonstrates that among ETGs with tidal features, those with shells exhibit a much lower $\lambda_{R_e}$ than those with tails or streams. The median $\lambda_{R_e}$ for ETGs with shells is $0.13\pm0.02$, while for those with tails and streams, the median $\lambda_{R_e}$ values are $0.22\pm0.02$ and $0.22\pm0.01$, respectively. KS tests reveal that the distributions of $\lambda_{R_e}$ are significantly different (with $p\sim6\times10^{-4}$) between ETGs with shells and those with streams or tails, whereas there is no significant difference in the $\lambda_{R_e}$ distributions (with $p=0.89$) between ETGs with streams and those with tails. However, the $\lambda_{R_e}$ values of ETGs with tails or streams are still significantly lower than those of ETGs without tidal features, given that $p$ is less than $0.01$.

The fractions $f_\mathrm{slow}$, calculated from the criterion of \citet{Cappellari2016}, are $0.46\pm0.06$, $0.20\pm0.05$, and $0.22\pm0.03$ for ETGs with shells, tails, and streams, respectively.\footnote{Based on the criterion of \citet{Emsellem2011}, the $f_\mathrm{slow}$ values are $0.43\pm0.06$, $0.19\pm0.05$, and $0.21\pm0.03$ for ETGs with shells, tails, and streams, respectively.} Thus, nearly half of ETGs with shells are slow rotators. This fraction is more than twice as high as that of ETGs with tails or streams, and more than three times higher than that of ETGs without tidal features.

The right panel of Figure \ref{fig:mass} displays the distribution of ETGs with shells, tails, and streams, as well as those without tidal features, in the $\lambda_{R_e}$ versus $\log M_\mathrm{star}$ plane. The upper and middle panels of Figure \ref{fig:bin_class} show the mean $\lambda_{R_e}$ and $f_\mathrm{slow}$, respectively, as a function of $\log M_\mathrm{star}$ and $\sigma_e$ for ETGs with shells, tails, streams, and those without tidal features. The lower panels of Figure \ref{fig:bin_class} display histograms of $\log M_\mathrm{star}$ and $\sigma_e$ for each ETG category. 

Figure \ref{fig:bin_class} reveals that ETGs with streams and those with tails exhibit very similar $\lambda_{R_e}$ and $f_\mathrm{slow}$ in each bin, with the differences in $\lambda_{R_e}$ and $f_\mathrm{slow}$ being less than 0.008. They also display comparable trends as a function of mass. Moreover, they have similar mass distributions, with median $\log(M_\mathrm{star}/M_\odot)$ values of 10.8. Due to their similarity, ETGs with streams and those with tails are grouped together in the subsequent description.

Figures \ref{fig:mass} and \ref{fig:bin_class} demonstrate that $\lambda_{R_e}$ of ETGs with streams + tails are lower by $\sim0.05$ than that of ETGs without tidal features in all the $M_\mathrm{star}$ and $\sigma_e$ bins. Furthermore, ETGs with shells have $\lambda_{R_e}$ lower by $\sim0.07$ than ETGs with streams + tails in all the $M_\mathrm{star}$ and $\sigma_e$ bins. KS tests on the distributions of $\lambda_{R_e}$ for ETGs with streams + tails  and those without tidal feature, corrected for the variation of $\lambda_{R_e}$ as a function of $M_\mathrm{star}$ and $\sigma_e$, give $p<5\times10^{-5}$. The same tests on the distributions of $\lambda_{R_e}$ for ETGs with shells and those with streams + tails yield $p<5\times10^{-3}$. These KS tests suggest that the difference in $\lambda_{R_e}$ between any two ETG groups remains significant even after accounting for the relation between $\lambda_{R_e}$ and mass.

These results are reflected in the trend for $f_\mathrm{slow}$ as displayed in the middle panels of Figure \ref{fig:bin_class}. In all the $M_\mathrm{star}$ and $\sigma_e$ bins, $f_\mathrm{slow}$ of ETGs with shells is higher by $\sim0.2$ than that of ETGs with streams + tails, while $f_\mathrm{slow}$ of ETGs with streams + tails is higher by $\sim0.04$ than that of ETGs without tidal features in the low-mass range.

We note that ETGs with purely shell-type tidal features and no features of other types have slightly lower $\lambda_{R_e}$ by $\sim0.01$ (resulting in a higher $f_\mathrm{slow}$ by $\sim0.04$) in each mass bin than ETGs in which at least one of the tidal features is a shell. By contrast, ETGs that have purely stream- or tail-type tidal features without shells exhibit very slightly higher $\lambda_{R_e}$ by $\sim0.01$ (hence a lower $f_\mathrm{slow}$ by $\sim0.02$) in each mass bin than ETGs in which at least one of the tidal features is a stream or a tail.\footnote{ETGs with pure streams or tails still have lower $\lambda_{R_e}$ than those without tidal features.}

The histograms shown in the lower panels of Figure \ref{fig:bin_class} show that shells are typically found in ETGs that are more massive by $0.2$ dex than those with streams and tails. The median $\log(M_\mathrm{star}/M_\odot)$ values are $11.01\pm0.05$ and $10.82\pm0.03$ for ETGs with shells and those with streams + tails, respectively. A KS test conducted on the distributions of $\log(M_\mathrm{star}/M_\odot)$ for the two ETG categories yields $p=8\times10^{-4}$, indicating a significant difference between the two distributions.

In sum, ETGs with tidal features have a lower $\lambda_{R_e}$ and hence a higher $f_\mathrm{slow}$ than ETGs without tidal features. When tidal features are subdivided into shells and streams/tails, ETGs with shells, which exhibit a much lower $\lambda_{R_e}$ than those with tails or streams, contribute the most to reducing $\lambda_{R_e}$ (increasing $f_\mathrm{slow}$), while the $\lambda_{R_e}$ values of ETGs with tails or streams are still lower than those of ETGs without tidal features. These findings generally remain valid even when ETGs are divided into several mass bins. In addition, tidal features are more frequently detected in more massive ETGs, with shells typically found in slightly more massive ETGs compared to streams and tails.
\\

\section{Discussion}\label{sec:discuss}

Our results imply that mergers that generate shell-type tidal features also lower $\lambda_{R_e}$ more effectively than mergers that create tidal streams or tails. The simulations in previous studies \citep{Moody2014,Lagos2018b,Li2018} show that radial infall of satellites (radial mergers with orbits of low-angular momentum) can reduce the stellar angular momentum (rotation support) of merger remnants more effectively than circular infall of satellites (mergers with orbits of high angular momentum). Another set of simulations \citep{Quinn1984,Dupraz1986,Johnston2008,Hendel2015,Pop2018,Karademir2019} suggests that radial mergers are capable of producing shell-type tidal features, whereas satellite infall with circular orbits can make stream-like linear tidal features. Therefore, our results support the scenario that radial mergers, which are better at reducing $\lambda_{R_e}$ compared to circular mergers, are more closely associated with the formation of shell-type tidal features rather than tidal streams or tails.

An alternative hypothesis that may explain our results is that the infall of galaxies into ETGs with already low $\lambda_{R_e}$ tends to result in the formation of shell-type tidal features. However, this should be verified through the analysis of galaxy simulations. Another potential explanation for our findings is that ETGs with streams could be, on average, at earlier stages of merging (hence $\lambda_{R_e}$ has not fully reduced yet) than those with shells because circular orbits have longer merger timescales than radial orbits.

The finding that tidal features are more frequently detected in more massive ETGs aligns well with the understanding that the formation of more massive ETGs involves a higher number of mergers \citep{Dubois2016,Yoon2017,Davison2020}. Our finding that shells are typically found in slightly more massive ETGs compared to streams and tails may be due to the nature of dynamical friction. Dynamical friction is stronger in the case when infall galaxies are more massive (major mergers). This is because massive satellites exert a greater influence on their surroundings and consequently experience stronger dynamical friction. Since dynamical friction can make the infall orbit of satellite galaxies more radial, massive satellites are more likely to be accreted through radial orbits in the later stages of mergers \citep{Amorisco2017,Pop2018}. This could explain our observational result that shell-type tidal features are detected in slightly more massive ETGs, which is also found in the simulation of \citet{Pop2018} and in the observation of \citet{Bilek2023}. We note that low-mass satellites require almost purely radial infall orbits at the initial accretion time in order to produce shells in the simulation of \citet{Pop2018}.

As stated in Section \ref{sec:tf} and illustrated in Figures \ref{fig:main} and \ref{fig:mass}, among the 254 ETGs with tidal features, 27 have both tails and streams, and 25 have both shells and streams. However, in contrast, only five ETGs exhibit both shells and tails, which is a notably smaller count. This can be explained if tails and shells predominantly stem from major or intermediate mergers, whereas streams, the most prevalent tidal features observed in 177 ETGs in our sample, typically arise from more frequent minor mergers. Indeed, many previous studies associate tails with major mergers and streams with minor mergers \citep{Duc2015,Mancillas2019,Bilek2020,Bilek2023,Sola2022}, and our findings indicate that shells are more likely formed through the accretion of massive satellites. Therefore, the simultaneous occurrence of shells and tails is rare, likely due to the low possibility of two major or intermediate mergers occurring consecutively within the lifespan of tidal features.

In addition, if the mass ratio of galaxy mergers is the decisive factor for distinguishing between the formation of tails and streams, our finding that ETGs with tails and those with streams exhibit identical distributions in $\lambda_{R_e}$ implies that the kinematics of merger remnants may not be significantly determined solely by the mass ratio of the merging galaxies, highlighting the importance of other factors influencing their kinematics.
\\

\section{Summary}\label{sec:summary}

We investigate differences in stellar kinematics ($\lambda_{R_e}$) among ETGs with several types of tidal features and those without tidal features. This is done by categorizing tidal features, which serve as direct evidence of recent mergers, into shells, streams, and tails. Through this study, we broaden our understanding of the impact of galaxy mergers on $\lambda_{R_e}$ of ETGs. This study is an extension of that of \citet{Yoon2022}, using a sample of ETGs that is more than six times larger, which enables us to divide tidal features into several categories.

We use MaNGA IFU data for the analysis of the stellar kinematics of ETGs. In order to reduce the seeing effect in stellar kinematics, we apply deconvolution to the MaNGA IFU data using the LR algorithm. The pPXF code, which conducts full spectrum fitting on galaxy spectra, is used to extract stellar velocities and velocity dispersions. We detect and categorize tidal features through a visual inspection of DESI Legacy Survey images, which provide sufficient depth for the study of tidal features. The final sample consists of 1244 ETGs with redshifts of $z<0.055$ and stellar masses of $M_\mathrm{star}\ge10^{9.65}\,M_{\odot}$. The main results of this study are as follows.

\begin{enumerate}
\item ETGs with tidal features typically have reduced $\lambda_{R_e}$ values that are lower by $0.12$ (hence a higher fraction of slow rotators by $0.13$) compared to ETGs without tidal features, showing a significant difference in $\lambda_{R_e}$ distributions between the two ETG categories.

\item ETGs with shells, which have $\lambda_{R_e}$ values lower by $0.1$ than ETGs with tails or streams, contribute the most to reducing $\lambda_{R_e}$. As a result, nearly half of ETGs with shells are classified as slow rotators. This fraction is more than twice as high as that of ETGs with tails or streams, and over three times higher than that of ETGs without tidal features.

\item The $\lambda_{R_e}$ values of ETGs with tails or streams are slightly lower than those without tidal features, while there is no significant difference in $\lambda_{R_e}$ between ETGs with streams and those with tails.

\item These trends generally remain valid even when ETGs are segmented into several mass bins.

\item Our findings support the idea that radial mergers of low-angular momentum orbits, which are more effective at reducing $\lambda_{R_e}$  than circular mergers, are more closely related to the formation of shell-type tidal features than streams or tails.

\item Shells tend to be found in slightly more massive ETGs compared to streams and tails. This may be accounted for by the fact that massive satellite galaxies are more likely to be accreted through radial orbits, due to the nature of dynamical friction. 
\end{enumerate}

Increasing the size of the galaxy sample using deep and large survey images enhances our understanding of the origins of tidal features and the impact of galaxy mergers, as demonstrated in this study. We expect that much deeper images from future large-scale surveys will further advance our comprehensive understanding of galaxy mergers and tidal features.\\

\begin{acknowledgments}
This research was supported by the Korea Astronomy and Space Science Institute under the R\&D program (Project No. 2024-1-831-00), supervised by the Ministry of Science and ICT.
\end{acknowledgments}



\end{document}